\documentclass[twocolumn]{article} 
\begin{document}

\onecolumn

\begin{center}

{\Large{\bf Reply to the claim by van Dokkum et al. for a galaxy not
containing dark matter.}}\\

\vspace{1.0cm}
{\bf Riccardo Scarpa}\footnote{Contact person: riccardo.scarpa@gtc.iac.es} \\
{\it IAC - Instituto de astrof\'isica de Canarias,\\
C/O Via Lactea, s/n E38205 - La Laguna, Tenerife, Espa\~na}\\

\vspace{0.5cm}

{\bf Xavier Hernandez}\\
{\it Instituto de Astronom\'ia,
Universidad Nacional Aut\'onoma de M\'exico,\\
AP 70-264, Distrito Federal 04510, M\'exico}\\

\vspace{0.5cm}

{\bf Ricardo Adan Cort\'es Martin}\\
{\it Instituto de Astronom\'ia,
Universidad Nacional Aut\'onoma de M\'exico,\\
AP 70-264, Distrito Federal 04510, M\'exico}\\

\vspace{0.5cm}

{\bf Renato Falomo}\\
{\it INAF - Istituto Nazionale di Astrofisica, Osservatorio
  Astronomico di Padova,\\ Vicolo dell’Osservatorio 5, I-35122 Padova, Italy}\\

\vspace{0.5cm}

{\bf Martin L\'opez-Corredoira} \\
{\it IAC - Instituto de astrof\'isica de Canarias,\\
C/O Via Lactea, s/n E38205 - La Laguna, Tenerife, Espa\~na}\\

\vspace{1.5cm}

(Date: May 13, 2018)\\

\vspace{1.5cm}
{\Large{\bf Abstract}}
\end{center}

The following is a short comment on the recent claim made by van Dokkum
and collaborators about the existence of a low surface brightness
galaxy, NGC1052-DF2, not containing dark matter. A discovery used by
the authors to both reject proposals of a failure of Newtonian
dynamics in the low acceleration regime (e.g., MOND), and prove the
dark matter hypothesis is correct. It is shown here that the claim in
untenable.

\nopagebreak
\twocolumn
A claim has been recently made that the low surface brightness galaxy
NGC1052-DF2 does not contain dark matter, while a huge amount of it was
expected.  The result is important to demonstrate that any claim of a
failure of Newtonian dynamics in the low acceleration regime (e.g.,
MOND) is wrong and, similarly, that the dark matter inferred in other
galaxies does really exist.

The "discovery" was presented in two papers. One appeared in Nature,
possibly the most prestigious scientific journal of all, with title "A
galaxy lacking dark matter" by van Dokkum et al. 2018, Nature, 555,
629 \footnote{https://arxiv.org/pdf/1803.10237.pdf}
and the other appeared in
the Astrophysical Journal, one of the most influential astronomical journals,
with title "An Enigmatic Population of Luminous Globular Clusters
in a Galaxy Lacking Dark Matter", van Dokkum et al., 2018, ApJ, 856,
30. \footnote{http://iopscience.iop.org/article/10.3847/2041-8213/aab60b/pdf}
With such references, the common reader will have no doubt about
the correctness and importance of the discovery and will see himself
compelled to abandon any alternative idea. 

Thus, let's have a look to what these authors have actually found.  The
galaxy NGC1052-DF2 is an extremely low surface brightness galaxy so,
at present, there is no way to directly measure its velocity
dispersion.  Therefore, its mass must be derived using some other
indicator. Van Dokkum and collaborators used a set of 10 globular
cluster supposedly gravitationally bound to the galaxy, to measure
their line of sight velocity dispersion. Further assuming the system
is virialized and isotropic, they derive from that the dynamical mass
of the galaxy under Newtonian expectations. The dynamical mass turns
out to be similar to the luminous mass, so they concluded the
galaxy has no dark matter at all. If confirmed, this would be the
first ever case of such a galaxy.

The first problem we see is that, beside the positional proximity in
the sky, there is no other evidence that the globular clusters used to
derive the velocity dispersion are physically bound to the
galaxy. Neither is there any evidence that the system is virialized
and isotropic. What if, because of anisotropy, the line of sight
velocity dispersion is not representative of the true value?

And what if the system is not virialized? Of course, if
not virialized, the velocity dispersion of the clusters is not
representative of the mass of the galaxy and the claim is invalidated.

Even if all the above assumptions hold, it could perfectly be
the case that the globular cluster system studied by van Dokkum
and collaborators forms a flattened
disk with substantial rotational support and is being observed close
to face-on, as the image of the galaxy itself might suggest. In the
above case, the velocity dispersion along the line of sight is merely
indicative of the vertical dynamics of such a disk, and hence would
naturally be expected to be significantly smaller than the full
dynamical value.

A series of objection to long to be discussed in such an important
paper, obviously.

The typical approach to figure out whether an assumption is correct, or
at least reasonable, is to compare its consequences with what is known.

The luminous mass of NGC1052-DF2 is $\sim 10^8$ solar masses. As far
as we know, galaxies this small {\it do not} host significant globular
cluster systems. Thus, as van Dokkum and collaborators themselves point out,
NGC1052-DF2 turns out to contain 1000 times more globular clusters
than expected. Suspicious.

Globular clusters observed in different galaxies are known to have
similar properties. In particular, their luminosity function varies
little form galaxy to galaxy and, for this reason, researchers usually
refer to it as {\it universal} luminosity function.

Well, in the case of NGC1052-DF2 the globular cluster luminosity
function turns out to be different from any other known. Far from
concluding their initial assumption was wrong, the authors went
on to claim they have demonstrated the luminosity function {\it is not}
universal!

Thus, according to van Dokkum and collaborators, this galaxy is unique
in at least three different aspects: it is the only galaxy know not to
contain dark matter, it is the only galaxy know to have such an
extremely large number of globular clusters, and it is the only galaxy
hosting a population of globular clusters not obeying to the {\it universal}
luminosity function. All this supported only by their own claim that
the selected clusters are physically related to the galaxy.

Obviously, one could stop here to dismiss the original claim. However,
let's go on and check how the velocity dispersion compares with the
expected value of 32 km/s observed in local galaxies of similar mass
\footnote{All quantities used here are taken directly from the papers by van Dokkkum and collaboratos.}.
The quoted velocity dispersion of NGC1052-DF2 is 3.2 km/s, as derived
from 10 velocities. Any scientist knows that doing statistic with such
a small number of measurements is dangerous, and not sufficient to
base an exceptional claim on. Nevertheless, the authors went on to
calculate the dispersion and instead of doing the strait forward
calculation, they used a biweight approach that basically kicked out
one cluster (as any one can see in fig 3b of their Nature paper).
They basically neglected the velocity of cluster 98, precisely the
value that disagrees with their claim!  Indeed, the velocity
dispersion including all values is 14.8 km/s, 4.6 times the value
quoted in the paper as the more probable.  van Dokkum and
collaborators do place an upper limit of 10.5 km/s to the velocity
dispersion, at 90\% confidence level. Again, this level of confidence
is too low, it correspond to less than a 2sigma deviation from
expectation.  In other words, being the true dispersion 32 km/s and
everything else correct, we have 10\% probability to get by pure chance
the quoted dispersion upper limit.  Not enough to support a claim as strong as
the one made by van Dokkum. Actually, not enough to support any claim.
The standard is to require at least a 5sigma deviation, or 99.999\%
probability.  Thus, at most, the data presented by van Dokkum et
al. might mildly support the claim that NGC1052-DF2 has less dark
matter than average, nothing more than that (but remember that all
this is based on the assumption of a physical connection between the
galaxy and the globular clusters, that in turn must be close to
isothermal).

Having said all the above, we cannot avoid to ask ourselves how could such a
result have been published at all, wondering how a similar
paper would have been received if the conclusion were the other way
around.  This points to the responsibility of journals that at present
adopt standards orders of magnitude lower to publish results favouring
the dark matter hypothesis compared to the ones required to papers
claiming the opposite. Our sad conclusion is that science cannot
progress this way. 

\end{document}